\documentclass[nohyperref]{article}

\usepackage{microtype}
\usepackage{graphicx}
\usepackage{booktabs} 

\usepackage{xcolor} 
\usepackage{times}
\usepackage{epsfig}
\usepackage{amsmath}
\usepackage{amssymb}

\usepackage[bb=dsserif]{mathalpha}
\usepackage{bm}

\usepackage{booktabs,colortbl,tabularx}
\usepackage{pifont}%

\usepackage[accepted]{icml2023}

\usepackage{amsmath}
\usepackage{amssymb}
\usepackage{mathtools}
\usepackage{amsthm}
\usepackage{microtype}

\usepackage{multirow}
\usepackage{comment} 
\usepackage{enumitem}

\usepackage{makecell}

\usepackage{bbm}

\usepackage{tikz}
\usepackage{comment}

\usepackage{algorithm}
\usepackage{algorithmic}

\usepackage{caption}
\usepackage{subcaption}

\theoremstyle{plain}

\theoremstyle{definition}

\theoremstyle{remark}

\definecolor{Gray}{gray}{0.9}
\definecolor{citecolor}{HTML}{2980b9}
\definecolor{linkcolor}{HTML}{c0392b}

\usepackage[pagebackref=true,breaklinks=true,colorlinks,bookmarks=false,citecolor=citecolor,linkcolor=linkcolor]{hyperref}

\usepackage[capitalize,noabbrev]{cleveref}
\crefformat{section}{Sec.~#2#1#3}
\crefformat{subsection}{Sec.~#2#1#3}
\crefformat{subsubsection}{Sec.~#2#1#3}
\crefformat{figure}{Fig.~#2#1#3}
\crefformat{equation}{Eq.~#2#1#3}
\crefformat{table}{Tab.~#2#1#3}
\crefformat{algorithm}{Alg.~#2#1#3}

\usepackage{xspace}
\newcommand{\method}{{\fontfamily{lmtt}\selectfont\emph{\textsc{OneAVM}}}\xspace}

\renewcommand{\v}{\mathbf{v}}
\renewcommand{\a}{\mathbf{a}}

\newlength\savewidth
\newlength\thinwidth

\definecolor{Gray}{gray}{0.93}

\usepackage{xparse}
\usepackage{pifont}
\usepackage{bm}
\usepackage{xspace} 
\makeatletter
\DeclareRobustCommand\onedot{\futurelet\@let@token\@onedot}
\def\@onedot{\ifx\@let@token.\else.\null\fi\xspace}
\def\eg{\emph{e.g}\onedot} 
\def\ie{\emph{i.e}\onedot}

\makeatother

\newcommand{\magenta}[1]{{\color{magenta} #1}}

\definecolor{battleshipgrey}{rgb}{0.52, 0.52, 0.51}
\renewcommand{\algorithmiccomment}[1]{\textcolor{battleshipgrey}{\#\ #1}}

\icmltitlerunning{A Unified Audio-Visual Learning Framework for Localization, Separation, and Recognition}

\begin{document}

\twocolumn[
\icmltitle{A Unified Audio-Visual Learning Framework for \\ Localization, Separation, and Recognition}

\begin{icmlauthorlist}
\icmlauthor{Shentong Mo}{CMU}
\icmlauthor{Pedro Morgado}{UWM} \\
\magenta{\url{https://github.com/stoneMo/OneAVM}}
\end{icmlauthorlist}

\icmlaffiliation{CMU}{Carnegie Mellon University}
\icmlaffiliation{UWM}{University of Wisconsin-Madison, Department of Electrical and Computer Eng}
\icmlcorrespondingauthor{Shentong Mo}{shentonm@andrew.cmu.edu}

\icmlkeywords{Audio-Visual Learning, Source Separation, Source Localization, Audio-Visual Recognition}
\vskip 0.3in
]



\printAffiliationsAndNotice{} 

\begin{abstract}

The ability to accurately recognize, localize and separate sound sources is fundamental to any audio-visual perception task. 
Historically, these abilities were tackled separately, with several methods developed independently for each task. However, given the interconnected nature of source localization, separation, and recognition, independent models are likely to yield suboptimal performance as they fail to capture the interdependence between these tasks.
To address this problem, we propose a unified audio-visual learning framework (dubbed \method) that integrates audio and visual cues for joint localization, separation, and recognition.
\method comprises a \textit{shared} audio-visual encoder and task-specific decoders trained with three objectives. The first objective aligns audio and visual representations through a localized audio-visual correspondence loss. The second tackles visual source separation using a traditional mix-and-separate framework. Finally, the third objective reinforces visual feature separation and localization by mixing images in pixel space and aligning their representations with those of all corresponding sound sources.
Extensive experiments on MUSIC, VGG-Instruments, VGG-Music, and VGGSound datasets demonstrate the effectiveness of \method for all three tasks, audio-visual source localization, separation, and nearest neighbor recognition, and empirically demonstrate a strong positive transfer between them. 
\end{abstract}

\section{Introduction}

Identifying, localizing, and recognizing objects or movements are critical cognitive functions, often requiring joint visual and auditory processing. 
In fact, a variety of neurophysiological studies have demonstrated how audio-visual interactions play a critical role in human perception. For example, humans can localize events more accurately and precisely in the presence of audio-visual sensory data compared to unisensory conditions~\cite{odegaard2015biases}. Cross-modal integration effects have also been identified as the cause of enhanced visual processing if reliably preceded by a sudden sound~\cite{frassinetti2002enhancement}, as well as enhanced audio-visual speech perception in noisy environments~\cite{schwartz2004seeing}. 
These studies provide substantial evidence in support of the benefits of joint audio-visual processing for a wide range of tasks, from recognition to localization and source separation.

\begin{figure}[t]
\centering
\includegraphics[width=\linewidth]{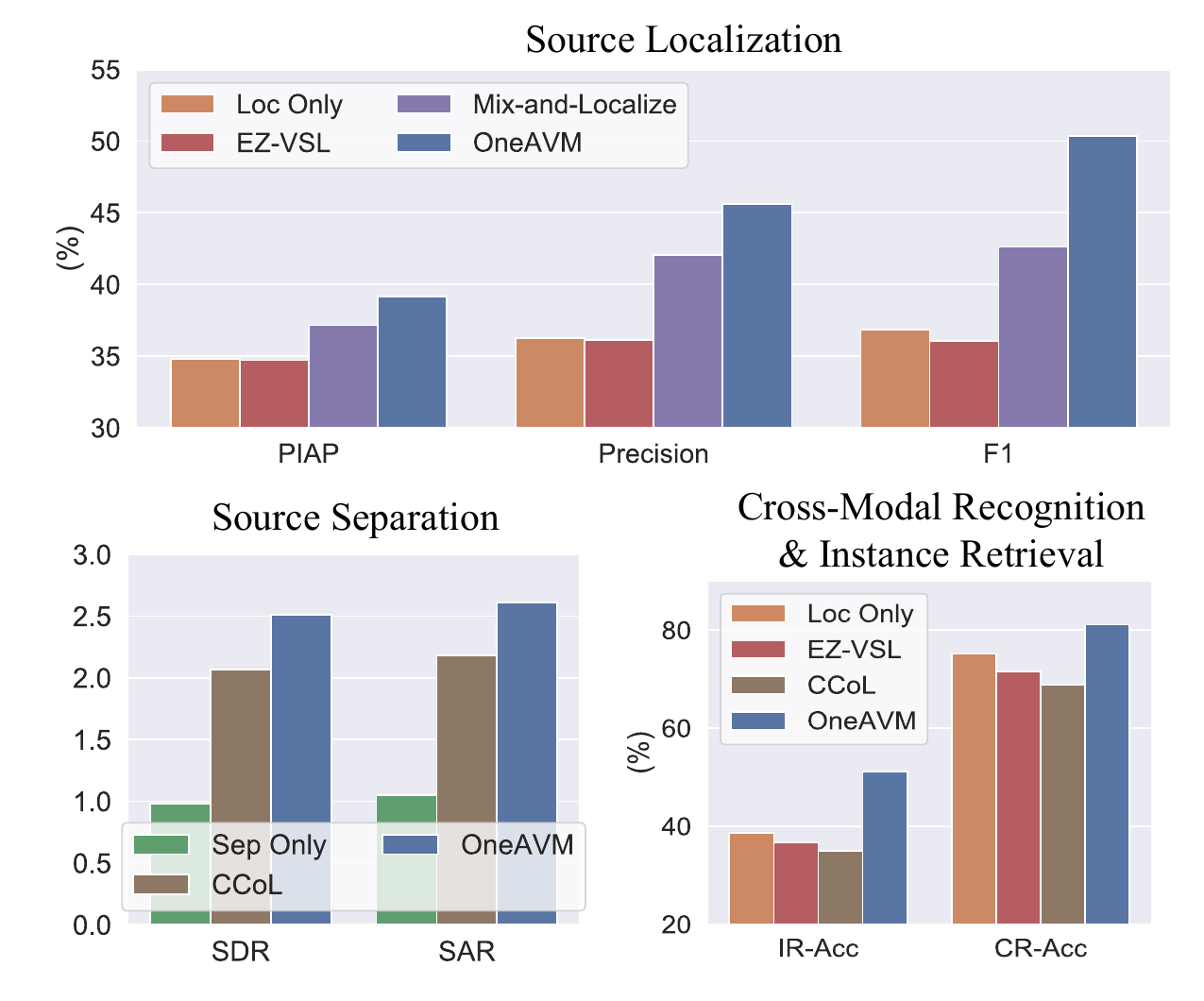}
\caption{Sound source localization, separation, recognition (CR-ACC), and instance retrieval (IR-ACC) performance of a variety of models. The proposed framework, not only surpasses prior state-of-the-art, but it does so using a single model without requiring task-specific finetuning.}
\label{fig: title_img}
\end{figure}

In the field of machine perception, researchers have also explored the potential of audio-visual interactions for each of these tasks separately. For instance, late multi-modal fusion was shown to greatly enhance object and action recognition~\cite{kazakos2019epic,xiao2020audiovisual}, particularly when the imbalance between dominant and non-dominant modalities is appropriately addressed. In a different line of work, two-stream neural networks, trained to match audio-visual representations, were shown capable of localizing sound sources in video data~\cite{Senocak2018learning,hu2022mix,mo2022EZVSL,mo2022SLAVC}, in a weakly supervised manner. Similarly, various multi-modal architectures have also been proposed to tackle the problem of sound source separation~\cite{hershey2001audio,zhao2018the,ephrat2018looking,Gao2018learning,xu2019mpnet,Gan2020music,tian2021cyclic}. 

Although promising, the aforementioned methods were developed independently for each audio-visual task and often rely on task-specific architectures and learning objectives. This not only limits the applicability of each model but also precludes positive transfer across tasks. In order to overcome these limitations and explore potential cross-task transfer, we propose a unified audio-visual learning framework capable of addressing several tasks simultaneously, including audio-visual source recognition, localization, and separation, without any task-specific fine-tuning. We dub our framework \method where "One" stands for the single unified model and AVM for audio-visual modeling.
We utilize a two-stream encoder shared across tasks with task-specific decoders. 
The model is trained to learn matched audio-visual representations, which enables it to identify cross-modal associations necessary for localization. Additionally, the model is required to produce representations conducive to effective source separation through a mix-and-separate objective. We also introduce a novel mixed visual alignment objective that aligns multiple audio representations to a corresponding mixture of images. Through these learning objectives, our framework provides a more comprehensive approach to audio-visual learning, enabling effective cross-task transfer and improving the applicability of audio-visual models.

Through extensive experiments on MUSIC, VGG-Instruments, VGG-Music, and VGGSound datasets, we show the cross-task transfer benefits obtained through our unified audio-visual framework, \method, as well as its state-of-the-art performance on visual sound localization, sound separation, and nearest neighbor recognition (see Figure~\ref{fig: title_img}). We highlight that, unlike the task-specific models of prior work, such capabilities are attained using a single model for all tasks. We further provide an extensive ablation study to validate the importance of simultaneous correspondence, localization, mixed audio separation, and mixed visual alignment in learning joint representations for the three downstream tasks.

In summary, this work provides three main contributions. (1) We present a novel unified audio-visual framework (\method) capable of performing sound source localization, separation, and recognition from a single model. (2) We propose a novel mixed visual alignment objective that associates individual sound sources with mixed images. (3) Extensive experiments comprehensively demonstrate the superiority of \method over previous baselines on various audio-visual downstream tasks.

\section{Related Work}

\noindent\textbf{Learning Representations from Audio-Visual Correspondences}.
Audio-visual representation learning aims to learn joint audio-visual models that can be tuned for various downstream tasks.
One extensive line of prior work ~\cite{aytar2016soundnet,owens2016ambient,Arandjelovic2017look,korbar2018cooperative,Senocak2018learning,zhao2018the,zhao2019the,Gan2020music,Morgado2020learning,Morgado2021robust,Morgado2021audio,hershey2001audio,ephrat2018looking,hu2019deep,mo2023diffava} 
have addressed audio-visual representation learning by establishing the correspondence between audio and visual modalities from videos. 
This cross-modal alignment has proved to be beneficial for several audio-visual tasks, including audio-visual spatialization~\cite{Morgado2018selfsupervised,gao20192.5D,Chen2020SoundSpacesAN,Morgado2020learning}, event localization~\cite{tian2018ave,lin2019dual,wu2019dual,lin2020audiovisual}, audio-visual navigation~\cite{Chen2020SoundSpacesAN,chen2021waypoints,chen22soundspaces2}, and parsing~\cite{tian2020avvp,wu2021explore,lin2021exploring,mo2022multimodal}.
In this work, we also learn from audio-visual correspondences as one of our objectives. However, we focus on integrating multiple tasks like sound source localization, separation, and recognition in a unified framework.

\begin{figure*}[t]
\centering
\includegraphics[width=1\linewidth]{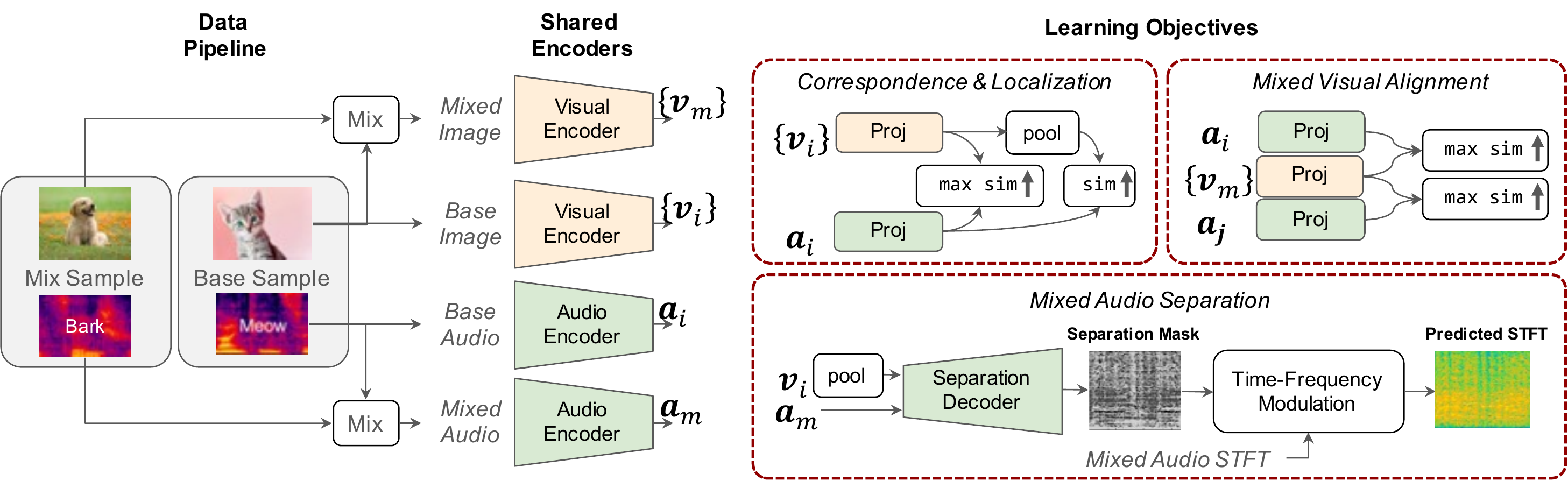}
\caption{Illustration of the proposed audio-visual learning framework for sound source localization, separation, and recognition.
First, image and audio encoders are applied to extract audio and visual features, which are trained for three separate objectives.
1) An audio-visual {\bf correspondence and localization} objective is utilized to align corresponding audio and visual features. 
2) An audio decoder is added for sound source separation using a mix-and-separate strategy ({\bf mixed audio separation}).
3) A novel {\bf mixed visual alignment} objective is proposed to align representations from a mixed image with the corresponding individual sound sources.
}
\label{fig: main_img}
\end{figure*}

\noindent\textbf{Visual Sound Source Localization}.
Visual sound source localization is a challenging task that seeks to identify objects or regions of a video corresponding to active sound sources.
Early works~\cite{hershey1999audio,fisher2000learning,kidron2005pixels} used conventional machine learning techniques, \eg, statistical models~\cite{fisher2000learning} and canonical correlation analysis~\cite{kidron2005pixels}, to learn low-level alignment between audio and visual features.
With the advance of deep neural nets, recent approaches~\cite{Senocak2018learning,hu2019deep,Afouras2020selfsupervised,qian2020multiple,chen2021localizing,arda2022learning,mo2022EZVSL,mo2022SLAVC,mo2023audiovisual,mo2023avsam} apply diverse neural-net based architectures to learn from audio-visual correspondences.
For example, Attention10k~\cite{Senocak2018learning} predicted regions of sounding objects in the image using a two-stream architecture with an attention mechanism.
To improve the localization performance, LVS~\cite{chen2021localizing} introduced hard sample mining to optimize a differentiable threshold-based contrastive objective for generating discriminative audio-visual correspondence maps.
Meanwhile, EZ-VSL~\cite{mo2022EZVSL} designed a multiple-instance contrastive learning loss to align regions with the most corresponding audio without negative regions involved.
More recently, a contrastive random walk framework was introduced in Mix-and-Localize~\cite{hu2022mix} to link each audio node with an image node using a transition probability of audio-visual similarity, which localizes individual sound sources from the mixture.
However, they do not involve reconstructing separated audio and audio-visual recognition during training.
Different from them, we aim to combine the sound localization objective with other audio-visual objectives, \textit{e.g.}, separation and recognition, in a unified framework to achieve general learning for the audio-visual community.

\noindent\textbf{Audio-Visual Source Separation}.
Audio-visual source separation aims at separating and recovering individual sound sources from an audio mixture given the image or an image region containing the sound source to be separated.
In recent years, several architectures and training frameworks have been proposed to enhance audio-visual source separation.~\cite{hershey2001audio,zhao2018the,ephrat2018looking,Gao2018learning,xu2019mpnet,tian2021cyclic,tzinis2020into}.
For instance, Zhao \textit{et al.}~\cite{zhao2018the} utilized the correspondence between sound sources and image regions for visually grounded separation.
Other approaches, such as MP-Net~\cite{xu2019mpnet} and CCoL~\cite{tian2021cyclic}, kept the mix-and-separate learning strategy of~\cite{zhao2018the} while improving the separation network architecture.
MP-Net~\cite{xu2019mpnet} applied a recursive MinusPlus Net to separate salient sounds from the mixture, and CCoL~\cite{tian2021cyclic} leveraged a cyclic co-learning framework which can benefit from the visual grounding of sound sources. Similarly to CCoL, our approach also jointly optimizes for separation and localization. We show, however, superior empirical performance compared to CCoL while utilizing a much simpler architecture (without complex detection heads).
Beyond visual appearance, other visual modalities have been recently explored to capture complicated visual representations, such as motion in SoM~\cite{zhao2019the}, gestures through pose and keypoint detection in MG~\cite{Gan2020mg}, and the spatiotemporal visual scene graphs in AVSGS~\cite{Chatterjee2021visual}. While promising, this work focuses on representations obtained from visual appearance alone (\ie, RGB frames). This choice enabled us to simplify the source separation architecture and focus on developing a unified framework for audio-visual learning. A promising avenue for future research is to explore how best to incorporate such visual modalities into a unified audio-visual model.

\noindent\textbf{Mixup Regularization Techniques.}
Mixup regularization techniques seek to create novel samples by combining existing training samples. In computer vision, popular techniques include Mix-Up~\cite{mixup} and CutMix~\cite{yun2019cutmix}, which randomly combine two or more samples or image patches to create new training examples, or Cutout regularization~\cite{devries2017cutout} which involves randomly removing pixels or patches from the input data to force the model to learn more robust features. These regularization techniques have proven effective in improving the robustness and generalization performance of deep learning models. We found them particularly useful in our unified audio-visual learning framework.  Thus, inspired by this line of work, we proposed a multi-modal variation of mix-up, where the representations of a mixed visual frame are trained to be aligned with the audio representations of all of the corresponding sound sources.

\section{Method}

The motivating hypothesis of this work is that audio-visual tasks like visual source localization, separation, and recognition can benefit from positive transfer across tasks. To achieve this goal, we created a unified learning framework and an audio-visual architecture to tackle these three tasks simultaneously. Our framework consists of three main components, namely correspondence \& localization, as explained in section~\ref{sec:bcl}, mixed audio separation in section~\ref{sec:mas}, and mixed visual alignment in section~\ref{sec:mva}.

\subsection{Problem Setup}
Our goal is to develop a unified model for audio-visual source localization, separation, and recognition, by training on an unlabeled audio-visual dataset $\mathcal{D}={(v_i, a_i): i=1, \ldots, N}$ leveraging the natural associations between audio and visual signals found in video data. 
The first task, audio-visual source localization, seeks to localize sound sources present in the audio $a_i$ within the visual frame $v_i$.
The second task, audio-visual source separation, is tackled through a mix-and-separate strategy~\cite{zhao2018the}. 
The final task is recognition and instance retrieval, which aims to develop semantic representations of the data, where cross-modal association within a sample and object-level clusters across samples are readily available. To evaluate recognition and cross-modal instance retrieval, we use cosine similarity between high-level features for either retrieval or nearest neighbor classification.

Despite the disparity between the three tasks, audio-visual interactions that are useful (and easily learned) from one task might benefit the others. Therefore, to enable positive transfer across tasks, we process all audio-visual pairs through a shared encoder, regardless of the target task. Specifically, we utilize a two-stream neural network, where audio representations $\a_i$ are generated by an audio encoder $f_a$, and localized visual representations $\v_i^{xy}$ are computed through a visual encoder $f_v$. Convolutional neural networks are used for both encoders, with ResNet-18 being the chosen architecture. The audio encoder input is a log-mel spectrogram. For the visual encoder, we obtain the localized representations $\v_i^{xy}$ from the last feature map before global pooling.
These latent representations are then fed to task-specific prediction heads, described in the following subsections.

\subsection{Correspondence \& Localization}~\label{sec:bcl}
Global audio-visual correspondence has been shown to lead to semantic aware representations useful for recognition tasks~\cite{Arandjelovic2017look,Morgado2021audio}. On the other hand, localization requires an objective that promotes spatially localized audio-visual correspondence~\cite{Senocak2018learning,mo2022EZVSL}. To accomplish both tasks, we seek both locally and globally aligned representations. 
Specifically, we apply two projection heads $\a_i^{\text{glb}}=g_a^{\text{glb}}(\a_i)$ and $\a_i^{\text{loc}}=g_a^{\text{loc}}(\a_i)$ to obtain two representations of audio $a_i$. Similarly, we project a globally pooled visual representation  $\v_i^{\text{glb}} = g_v^{\text{glb}}(\max_{xy} \v_i^{xy})$, as well as a set of local visual representations $V_i^{\text{loc}} = \{g_v^{\text{loc}}(\v_i^{xy}): \forall x,y\}$ of the base visual frame $v_i$. 
To align representations at both the local and global level, we define the audio-visual correspondence score between an audio $a_i$ and video $v_j$ as 
\begin{equation}
    s_{ij} = \mathtt{sim}(\mathbf{a}_i^{\text{glb}}, \mathbf{v}_j^{\text{glb}}) + \max_{\v_j^{loc}\in V_j^{loc}} \mathtt{sim}(\a_i^{\text{loc}}, \v_j^{loc}),
\end{equation}
where $\mathtt{sim}(\cdot, \cdot)$ represents a cosine similarity between the audio and visual features. The model is then trained to optimize the average cross-modal instance discrimination loss defined as
\begin{equation}\label{eq:cor}
\begin{aligned}
    \mathcal{L}^{\text{CL}}_i = 
    & - \log \frac{
    \exp \left( s_{ii} / \tau \right)
    }{
    \sum_{j=1}^B \exp \left(  s_{ij} / \tau \right)} - \log \frac{
    \exp \left( s_{ii} / \tau \right)
    }{
    \sum_{k=1}^B \exp \left( s_{ki} / \tau\right)},
\end{aligned}
\end{equation}
where $\tau$ is a temperature hyper-parameter, and the $B-1$ negatives are other samples in the current batch.
It is worth noting that this formulation would be equivalent to the audio-visual correspondence objective of~\cite{Morgado2021robust} if the similarity was computed only on global representations $s_{ij}=\mathtt{sim}(\mathbf{a}_i^{\text{glb}}, \mathbf{v}_j^{\text{glb}})$, and would be equivalent to the multiple instance contrastive learning framework of~\cite{mo2022EZVSL}, if the audio-visual similarity was obtained from local representations alone, i.e.~${s_{ij}=\max_{\v_j^{loc}\in V_j^{loc}} \mathtt{sim}(\a_i^{\text{loc}}, \v_j^{loc})}$.

\subsection{Mixed Audio Separation}~\label{sec:mas}
In addition to correspondence and localization, our unified framework also tackles the cocktail party source separation problem using a mix-and-separate learning framework~\cite{zhao2018the}.
This involves randomly selecting two samples, $(a_i, v_i)$ and $(a_j, v_j)$, from the training set to create a mixed audio waveform $a_m=a_i+a_j$. An audio U-Net decoder $g_\text{sep}$ is then trained to recover the waveform $a_i$ from the audio mixture $a_m$, given the corresponding visual frame $v_i$. Specifically, the decoder receives the representations of the audio mixture $\a_{m}$ and the visual embeddings of the base sample $\v_i$, and applies a series of transposed convolutions and an output head to predict a time-frequency separation mask $\hat{M}_i=g_\text{sep}(\a_{m}, \v_i)\in\mathbb{R}^{T \times F}$. 
This separation mask is then used to modulate the input mixture STFT to separate the base audio.
\begin{equation}
    \hat{a}_i = \textit{iSTFT}\left( \textit{STFT}(a_m) \cdot \hat{M}_i \right)
\end{equation}
Similarly to~\cite{zhao2018the}, the target masks $M_i$ indicate the time-frequency bins in which the source is the most dominant component in the mixture, $M_i = \mathbb{1}_{|\textit{STFT}(a_i)|>|\textit{STFT}(a_m)|}\in\{0,1\}^{T \times F}$ where $\mathbb{1}$ is the indicator function applied to each \textit{STFT} bin separately.
Source separation is achieved by optimizing a binary cross-entropy loss over these binary targets $M_i$
\begin{equation}
    \mathcal{L}_i^{\text{MAS}} = \frac{1}{TF}\sum_{t=1}^T\sum_{f=1}^F\mbox{BCE}\left(\hat{M}_i(t, f), M_i(t, f)\right).
\end{equation}

\begin{table*}[t]
	\renewcommand\tabcolsep{6.0pt}
	\centering
 \caption{{\bf Sound source localization.} Quantitative results on VGGSound-Instruments, VGGSound-Music, VGGSound-All.}
   \label{tab: exp_sota_loc}
	\scalebox{0.8}{
		\begin{tabular}{l|ccc|ccc|ccc}
			\toprule
			\multirow{2}{*}{Method} & \multicolumn{3}{c|}{VGGSound-Instruments} & \multicolumn{3}{c|}{VGGSound-Music}  & \multicolumn{3}{c}{VGGSound-All} \\
			& PIAP(\%) & Precision(\%) & F1(\%) &  PIAP(\%) & Precision(\%) & F1(\%) &  PIAP(\%) & Precision(\%) & F1(\%) \\ 	
			\midrule
			Attention10k & 41.25 & 28.32 & 33.67 & 18.65 & 23.97 & 16.75 & 15.32 & 19.21 & 13.12 \\

               OTS & 47.51 & 25.71 & 29.85 & 25.52 & 27.52 & 26.16 & 29.82 & 32.82 & 25.87 \\

		   DMC & 45.32 & 26.52 & 30.37 & 24.37 & 25.73 & 18.06 & 20.16 & 23.90 & 16.37 \\

            CoarsetoFine & 40.22 & 27.23 & 32.09 & 26.19 & 28.73 & 27.58 & 28.21 & 29.13 & 21.53 \\

            DSOL & 47.85 & 50.22 & 52.15 & 37.26 & 42.51 & 43.08 & 30.56 & 35.72 & 29.01 \\

			LVS & 42.33 & 32.61 & 45.72 & 32.05 & 33.67 & 32.53 & 29.62 & 34.43 & 27.53 \\

			EZ-VSL & 43.80 & 38.53 & 52.36 & 34.72 & 36.15 & 36.07 & 31.33 & 37.79 & 31.32 \\

                Mix-and-Localize  & 47.32 & 49.73 & 58.75 & 37.15 & 42.07 & 42.62 & 32.31 & 36.35 & 32.15 \\

                \method (ours) & \textbf{50.67} & \textbf{55.21} & \textbf{67.26} & \textbf{39.16} & \textbf{45.63} & \textbf{50.37} & \textbf{34.52} & \textbf{39.68} & \textbf{38.75} \\
			\bottomrule
			\end{tabular}}
\end{table*}

\subsection{Mixed Visual Alignment}
~\label{sec:mva}
Inspired by mixup~\cite{mixup}, we introduce a regularization technique called {\it mixed visual alignment} (MVA). Specifically, we mix visual frames from two samples, $v_i$ and $v_j$, with a mix-up coefficient $\alpha$ to produce a mixed frame $v_m = \alpha \cdot v_i + (1-\alpha) \cdot v_j$. Then, we align the visual representation of the mixed frame $v_m$ with the representations of both audios $a_i$ and $a_j$.
Let $\a_i^{\text{mva}}=g_a^{\text{mva}}(\a_i)$ and $\a_j^{\text{mva}}=g_a^{\text{mva}}(\a_j)$ be the two audio representations, and $V_{m}^{\text{mva}}=\{g_v^{\text{mva}}(\v_m^{xy}): \forall x,y\}$ the set of local visual features for mixed frame $v_m$, obtained through audio and visual projection heads, $g_a^{\text{mva}}(\cdot)$ and $g_v^{\text{mva}}(\cdot)$, respectively. 
Mixed visual alignment is obtained by minimizing 
\begin{equation}\label{eq:mva}
\begin{aligned}
    \mathcal{L}^{\text{MVA}}_i = 
    & \alpha \mathcal{L}^{\text{CL}}(V_m^\text{MVA}, \a_i^\text{MVA}) + (1-\alpha) \mathcal{L}^{\text{CL}}(V_m^\text{MVA}, \a_j^\text{MVA})
\end{aligned}  
\end{equation}
where $\alpha$ is the mix-up coefficient.
The overall objective of our model is optimized in an end-to-end manner as:
\begin{equation}\label{eq:all}
    \mathcal{L} = \dfrac{1}{N}\sum_{i=1}^N (\mathcal{L}^{\text{CL}}_i + \mathcal{L}^{\text{MAS}}_i +\mathcal{L}^{\text{MVA}}_i)
\end{equation}
We did not find it necessary to
add weighing constants to the loss terms for improving downstream performance.

\section{Experiments}

\subsection{Experimental setup}

\noindent\textbf{Datasets.}
We conducted experiments on the following audio-visual datasets.
1) {\it MUSIC}~\cite{zhao2018the} consists of 448 untrimmed YouTube music videos of solos and duets from 11 instrument categories. We use 358 solo videos for training and 90 solo videos for evaluation. Since some videos are no longer publicly available, the used dataset is slightly smaller than the original MUSIC dataset.
For a fair comparison, we trained all models (including prior work) on the same training data.
2) {\it VGGSound-Instruments}~\cite{hu2022mix} is a subset of VGG-Sound~\cite{chen2020vggsound} which includes 32k video clips of 10s lengths from 37 musical instruments categories for training and 446 videos for testing. Each video only has a single instrument category label. 
3) We composed another more challenging musical subset from VGG-Sound~\cite{chen2020vggsound} containing 40,908 video clips from 49 music categories for training and 1201 clips for testing. We refer to this subset {\it VGGSound-Music}.
4) Beyond the musical datasets, we used 150k video clips from 221 categories in VGG-Sound~\cite{chen2020vggsound}, denoted as {\it VGGSound-All}, where 221 classes are available in VGG-Sound Sources with source localization annotations.
For testing, we used the full VGG-Sound Source~\cite{chen2021localizing} test set, which contains 5158 videos with source localization annotations.
5) We also used the Kinetics-400 dataset~\cite{carreira2017kinetics} to demonstrate the benefits of pre-training. Kinetics contains 187k video clips of human actions across 400 categories.

\noindent\textbf{Evaluation Metrics.}
Following the prior work~\cite{hu2022mix,mo2022EZVSL,mo2022SLAVC}, we use the pixel-wise average precision (PIAP) from~\cite{hu2022mix}, as well as the Precision and F1 scores defined in ~\cite{mo2022SLAVC} for visual source localization.
For source separation, following ~\cite{zhao2018the}, we use Signal-to-Distortion Ratio (SDR) and Signal-to-Artifact Ratio (SAR).
Recognition and retrieval evaluations are based on nearest-neighbors retrievals, using cosine similarity between the representations obtained from the shared encoders. We assess the accuracy of a cross-modal instance retrieval task, denoted IR-Acc, which determines how often the audio of a sample can be accurately retrieved from its visual component and vice-versa. We also assess the accuracy of both within and cross-modal nearest neighbor classifier, denoted wNN-Acc and xNN-Acc, which measures the class consistency across neighboring samples.

\noindent\textbf{Implementation.}
The input images are resized into a $224 \times 224$ resolution.
The audio is represented by log spectrograms extracted from $3s$ of audio at a sample rate of $8000$Hz. 
We follow the prior work~\cite{mo2022EZVSL} and apply STFT to generate an input tensor of size $128 \times 128$ ($128$ frequency bands over $128$ timesteps) using 50ms windows with a hop size of 25ms.
For the audio and visual encoder, we use the ResNet18~\cite{he2016resnet} to extract unimodal features and initialize the visual model using weights pre-trained on ImageNet~\cite{imagenet_cvpr09}.
Unless other specified, the decoder depth for mixed audio separation was set to 8, and the mixing coefficient for mixed visual alignment was set to $\alpha=0.5$.
For projection heads, we use one linear layer for each modality and each separate objective.  
The models were trained for 20 epochs using the Adam optimizer~\cite{kingma2014adam} with a learning rate of $1e-4$ and a batch size of $128$.

\subsection{Comparison to prior work}
We begin by comparing our unified model \method to prior work on audio-visual sound source localization, separation, and recognition.

\noindent {\bf Sound source localization.} 
To validate the effectiveness of the proposed \method on sound source localization, we compare to the following prior work:
1) Attention 10k~\cite{Senocak2018learning}  (CVPR'2018): the first baseline on sound source localization using a two-stream and attention-based neural net;
2) OTS~\cite{Arandjelovic2018ots} (ECCV'2018): a correspondence-based baseline for localization;
3) DMC~\cite{hu2019deep} (CVPR'2019): a deep multi-modal clustering approach based on audio-visual co-occurrences;
4) CoarsetoFine~\cite{qian2020multiple} (ECCV'2020): a two-stage approach using coarse-to-fine embeddings alignment;
5) DSOL~\cite{hu2020dsol} (NeurIPS'2020): a class-based method with two-stage training;
6) LVS~\cite{chen2021localizing} (CVPR'2021): a contrastive learning framework with hard negative mining to learn audio-visual correspondence maps;
7) EZ-VSL~\cite{mo2022EZVSL} (ECCV'2022): a recent weakly supervised localization framework based on multiple-instance contrastive learning;
8) Mix-and-Localize~\cite{hu2022mix} (CVPR'2022): a recent method based on a contrastive random walk on a graph of images and separated sound sources.

\begin{table}[t]
	\renewcommand\tabcolsep{6.0pt}
	\centering
 \caption{{\bf Sound source separation.} Quantitative results on MUSIC and VGGSound-Music datasets.}
   \label{tab: exp_sota_sep}
	\scalebox{0.7}{
		\begin{tabular}{l|cc|cc}
			\toprule
			\multirow{2}{*}{Method} & \multicolumn{2}{c|}{MUSIC} & \multicolumn{2}{c}{VGGSound-Music}  \\
			& SDR & SAR & SDR & SAR \\ 	
			\midrule
                NMF & -0.62	& 2.41	 &	-7.12	& -9.01 \\
                RPCA & 0.86 	& 3.81	 &	-5.53	& -7.82 \\
                Sound-of-Pixels & 4.55 	& 10.24 &		0.95	& 1.03 \\
                
                MP-Net & 4.82 	& \textbf{10.56} &		1.37	& 1.39 \\
                CCoL & 6.35 	& 9.75	 &	2.07 & 2.18 \\
                \method (ours) & \textbf{7.38} & 7.48 & \textbf{2.51} & \textbf{2.61} \\
			\bottomrule
			\end{tabular}}
\end{table}

Table~\ref{tab: exp_sota_loc} presents the source localization performance on VGGSound-Instruments, VGGSound-Music, and VGGSound-All datasets.
The proposed \method outperformed prior work on all metrics across all three datasets.
We achieve significant improvements over DSOL~\cite{hu2020dsol}, a class-supervised approach, as well as EZ-VSL~\cite{mo2022EZVSL} and Mix-and-Localize~\cite{hu2022mix}, two state-of-the-art weakly-supervised source localization methods. 
For example, on VGGSound-Instruments, we outperform the second-best method (DSOL) by 2.82 PIAP, 4.99 Precision, and 15.11 F1 score. On VGGSound-Music, the second-best method (also DSOL) was outperformed by 3.96 PIAP, 3.96 Precision, and 9.74 F1 score. Finally, the second-best method on VGGSound-All (Mix-and-Localize) was also outperformed by significant margins, 2.21 PIAP, 3.33 Precision, and 6.6 F1 score. 
These improvements demonstrate the effectiveness of unifying multiple tasks to learn better representations for visual sound source localization.

\begin{table}[t]
	\renewcommand\tabcolsep{3.5pt}
	\centering
 \caption{{\bf Nearest-neighbor recognition.} Quantitative results on VGGSound-Music dataset.}
   \label{tab: exp_sota_recg}
	\scalebox{0.9}{
		\begin{tabular}{lcccccc}
			\toprule
			Method & IR-Acc & xNN-Acc & wNN-Acc \\	
			\midrule
                Sound-of-Pixels & 26.92 & 59.09 & 43.64 \\
                MP-Net & 30.71 & 64.21 & 49.64 \\
                CCoL & 35.00 & 68.92 & 54.20 \\
                EZ-VSL & 36.69 & 71.56 & 56.10 \\
                Mix-and-Localize & 38.54 & 75.42 & 60.18 \\
                \method (ours) & \textbf{51.17} & \textbf{81.10} &  \textbf{60.91} \\
			\bottomrule
			\end{tabular}}
\end{table}

\begin{figure*}[t]
\centering
\includegraphics[width=0.9\linewidth]{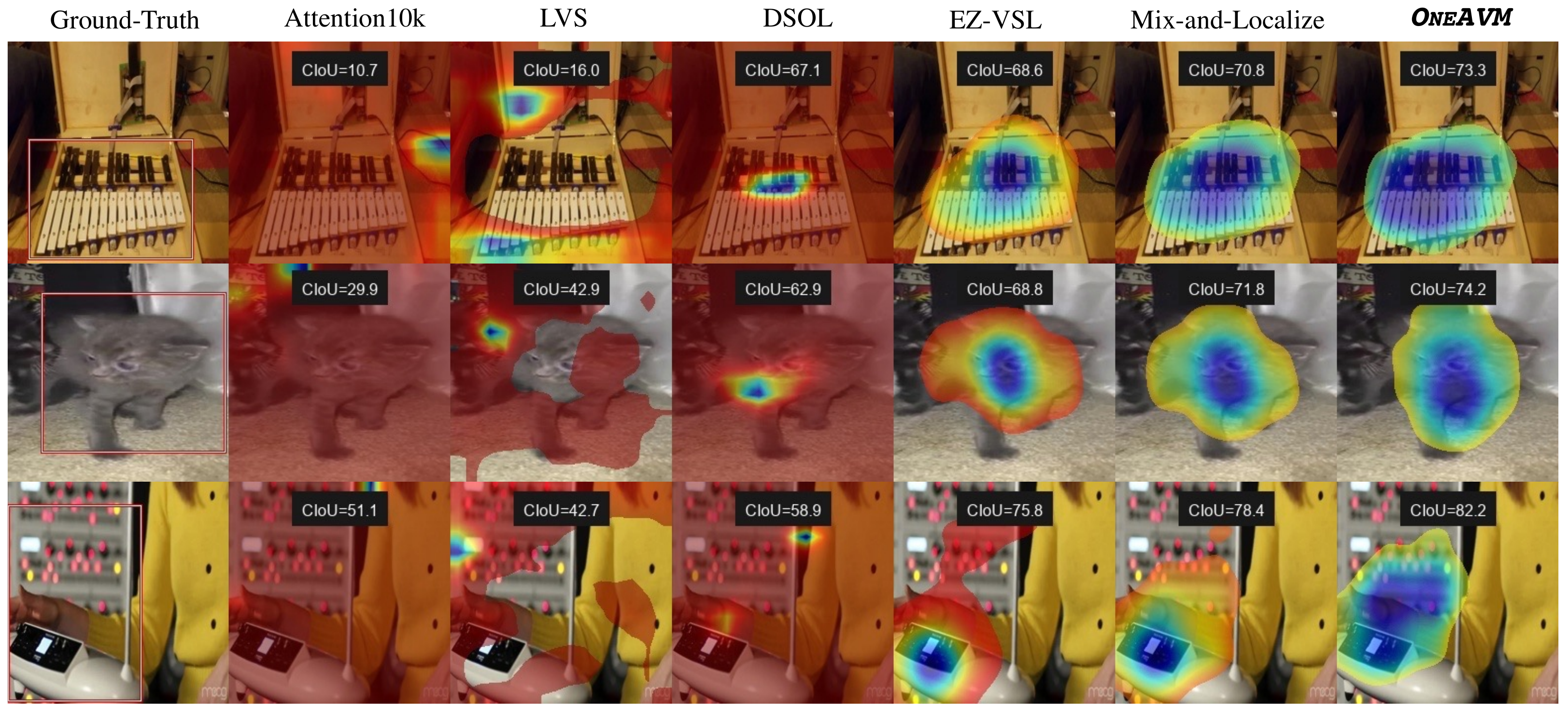}
\caption{Example sound source localization maps.
\method produces higher-quality localization maps for each source.
}
\label{fig: vis_loc}
\end{figure*}

\begin{table*}[!htb]
    \renewcommand\tabcolsep{6.0pt}
    \centering
    \caption{{\bf General video pre-training on Kinetics.} Source localization, separation, and nearest neighbor recognition performance on VGGSound-Music with and without Kinetics-400 pre-training.}
    \label{tab: exp_kinetics}
    \scalebox{0.8}{
    \begin{tabular}{ll|ccc|cc|ccc}
        \toprule
            \multirow{2}{*}{\bf Train DB} & \multirow{2}{*}{\bf Test DB} & \multicolumn{3}{c|}{\bf Localization} & \multicolumn{2}{c|}{\bf Separation} & \multicolumn{3}{c}{\bf Recognition}\\
          &  & PIAP & Precision & F1 &  SDR & SAR &  IR-Acc & xNN-Acc & wNN-Acc \\
        \midrule
        VGGSound-Music & VGGSound-Music & 39.16 & 45.63 & 50.37 & 2.51 & 2.61 & 51.17 & 81.10 & 60.91 \\
            Kinetics & VGGSound-Music & 22.89 & 26.91 & 38.78 & 0.65 & 0.71 & 8.82 & 9.61 & 22.81 \\
            Kinetics $\rightarrow$ VGGSound-Music & VGGSound-Music & \textbf{41.56} & \textbf{47.82} & \textbf{58.96} & \textbf{2.91} & \textbf{2.96} & \textbf{56.65} & \textbf{85.83} & \textbf{69.06} \\
        \bottomrule
    \end{tabular}}
\end{table*}

\noindent {\bf Sound source separation.} 
For source separation, we compare against the following methods:
1) NMF~\cite{Virtanen2007monaural}: a traditional signal processing approach based on non-negative matrix factorization to generate the spectrogram of each sound source;
2) RPCA~\cite{huang2012rpca}: a parameter-free baseline based on robust principal component analysis;
3) Sound-of-Pixels~\cite{zhao2018the}: a deep learning approach that recovers separated audio conditioned on pixel-level visual features;
4) MP-Net~\cite{xu2019mpnet}: an improved audio-visual method based on recursive separation from the mixture;
5) CCoL~\cite{tian2021cyclic}: a cyclic co-learning framework based on sounding object visual grounding to separate individual sound sources.

The comparison is shown in Table~\ref{tab: exp_sota_sep} on two datasets, MUSIC and VGGSound-Music. On the small MUSIC benchmark, we observe mixed results, with \method outperforming all prior work by more than 1.33 SDR, while achieving a SAR score lower than other source separation methods like Sound-of-Pixels, MP-Net, and CCoL. However, on the more challenging VGGSound-Music dataset, the proposed approach outperformed all prior work both in terms of SDR and SAR. In particular, \method outperforms methods that do not perform localization by significant margins (\eg outperforming MP-Net by 1.14 SDR and 1.22 SAR) and improves over CCoL, the only other method that benefits from joint localization and source separation.

\noindent {\bf Nearest-neighbor recognition.} 
We also compared \method with prior work on nearest-neighbor recognition and retrieval tasks, including Sound-of-Pixels~\cite{zhao2018the}, MP-Net~\cite{xu2019mpnet}, CCoL~\cite{tian2021cyclic}, EZ-VSL~\cite{mo2022EZVSL}, and Mix-and-Localize~\cite{hu2022mix}. Table~\ref{tab: exp_sota_recg} shows the comparison on the VGGSound-Music dataset. The proposed approach, \method, achieved the best performance across all metrics, outperforming the state-of-the-art localization methods like EZ-VSL and Mix-and-Localize by more than 12.6 points on cross-modal instance retrieval accuracy (IR-Acc), 5.7 points on cross-modal nearest neighbor accuracy (xNN-Acc) and 0.7 points on within-modal nearest neighbor accuracy (wNN-Acc).
On the other hand, prior separation methods like CCoL, MP-Net, and Sound-of-Pixels tend to underperform in recognition tasks compared to localization methods. This outcome is not unexpected as class information is not always learnable when training exclusively for source separation. Despite its occasional usefulness, class information is not a top priority for source separation.
\method can, however, achieve superior performance on all three tasks (separation, localization, and recognition) using a single model. In fact, all results on the VGGSound-Music dataset in Tables~\ref{tab: exp_sota_loc}, \ref{tab: exp_sota_sep} and \ref{tab: exp_sota_recg} were produced from the same model.

\begin{figure*}[t]
\centering
\includegraphics[width=0.8\linewidth]{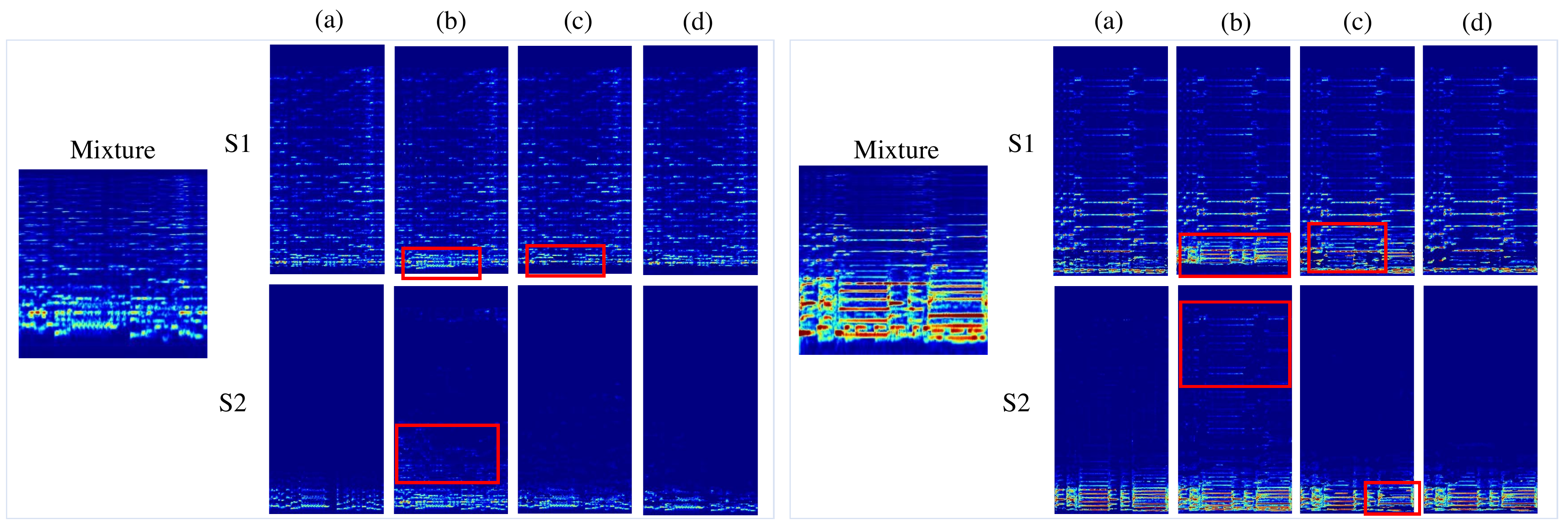}
\caption{Qualitative visualization of sound source separation.
(a) Ground-Truth; (b) Sound-of-Pixels; (c) MP-Net; (d) \method.
The proposed \method separates each source more accurately.
}
\label{fig: vis_sep}
\end{figure*}

\newcommand{\vsl}{
\begin{tabular}{cccccc}
    \toprule
    \bf CL & \bf MAS & \bf MVA & \bf PIAP & \bf Precision & \bf F1  \\
    \midrule
	\checkmark & & & 34.82 & 36.27 & 36.85 \\
 \checkmark &  \checkmark &  & 36.51 & 39.03 & 40.72 \\
 \checkmark &  & \checkmark & 37.05 & 40.38 & 42.86 \\
 \checkmark & \checkmark & \checkmark & \textbf{39.16} & \textbf{45.63} & \textbf{50.37} \\
    \bottomrule
\end{tabular}
}

\newcommand{\avss}{
\begin{tabular}{ccccc}
    \toprule
    \bf CL & \bf MAS & \bf MVA & \bf SDR & \bf SAR  \\
    \midrule
	& \checkmark &  & 0.98 & 1.05 \\
 \checkmark &  \checkmark & & 1.57 & 1.63 \\
 & \checkmark  & \checkmark & 1.98 & 2.08 \\
 \checkmark & \checkmark & \checkmark & \textbf{2.51} & \textbf{2.61} \\
    \bottomrule
\end{tabular}
}

\newcommand{\avrec}{
\begin{tabular}{cccccccccc}
    \toprule
    \bf CL & \bf MAS & \bf MVA & \bf AV-COR & \bf VA-COR & \bf AV-CLS & \bf VA-CLS & \bf AA-CLS & \bf VV-CLS  \\
    \midrule
	\checkmark &  & & 37.64	& 39.72	& 75.85 & 74.44 & 60.95 & 58.03 \\

 \checkmark &  \checkmark &  & 38.47	& 39.13	& 77.94 & 75.27 & 61.37 & 60.12 \\

 \checkmark &  & \checkmark & 43.56	& 45.38	& 78.82 & 76.03 & 61.12 & 59.96 \\

 \checkmark & \checkmark & \checkmark & \textbf{50.54} & \textbf{51.79} & \textbf{82.01} & \textbf{80.18} & \textbf{61.70} & \textbf{60.12} \\
    \bottomrule
\end{tabular}
}

\newcommand{\avrectwo}{
\begin{tabular}{ccccccc}
    \toprule
    \bf CL & \bf MAS & \bf MVA & \bf xIR-Acc & \bf xNN-Acc & \bf wNN-Acc  \\
    \midrule
    \checkmark &  & & 38.68	& 75.14 & 59.49 \\
    \checkmark &  \checkmark &  & 38.8	& 76.60 & 60.75 \\
    \checkmark &  & \checkmark & 44.47	& 77.43 & 60.54 \\
    \checkmark & \checkmark & \checkmark & \textbf{51.16} & \textbf{81.10}& \textbf{60.91} \\
    \bottomrule
\end{tabular}
}

\begin{table*}[t!]
    \centering
    \caption{{\bf Ablation studies.} Impact of Correspondence \& Localization (CL), Mixed Audio Separation (MAS), and Mixed Visual Alignment (MVA) for visual sound localization, sound source separation, and nearest-neighbor recognition. 
    \label{tab:ab_module}}
    \begin{subfigure}{0.34\textwidth}
        \resizebox{\linewidth}{!}{\vsl}
        \caption{Visual sound localization.}
        \label{tab:vsl}
    \end{subfigure}\hfill
    \begin{subfigure}{0.25\textwidth}
        \resizebox{\linewidth}{!}{\avss}
        \caption{Sound source separation.}
        \label{tab:avss}
    \end{subfigure}\hfill
    \begin{subfigure}{0.39\textwidth}
        \resizebox{\linewidth}{!}{\avrectwo}
        \caption{Nearest-neighbor recognition.}
        \label{tab:avrec}
    \end{subfigure}
\end{table*}

\noindent\textbf{General video pre-training.}
All datasets considered above, either based on VGGSound or MUSIC, are composed of video samples with relatively clean audio-visual associations. However, uncurated videos tend to display weaker associations.
To study the effect of uncurated videos on \method, we utilized Kinetics-400 (which was not composed to study audio events) for pre-training and transferred the learned model to VGGSound-Music by finetuning. Source localization, separation, and recognition performance were measured on VGGSound-Music. We set the pre-training schedule to 20 epochs, and the fine-tuning schedule to 10 epochs, both with a batch size of 128.

Table~\ref{tab: exp_kinetics} shows a comparison between 3 models: 1) trained on VGGSound-Music and evaluated on VGGSound-Music; 2) pre-trained on Kinetics and evaluated on VGGSound-Music; 3) pre-trained on Kinetics, finetuned on VGGSound-Music and then evaluated on VGGSound-Music.
Unsurprisingly, the model trained on Kinetics alone performs worse on every task. This can be due to the domain gap between Kinetics and VGG-Sound or the weaker audio-visual associations in Kinetics videos. 
Nevertheless, despite the noisier samples, pre-training on Kinetics still provides a strong initialization for transfer learning. This shows that general video data, which can be more easily collected at scale, still play an important role in pre-training unified audio-visual models.

\noindent {\bf Qualitative comparisons.} 
To further assess our unified framework, we show the localization maps and separated spectrograms generated by a single \method model in Figures~\ref{fig: vis_loc} and \ref{fig: vis_sep}, respectively, and compare them with the predictions produced by specialized methods, such Attention10k~\cite{Senocak2018learning}, LVS~\cite{chen2021localizing}, DSOL~\cite{hu2020dsol}, EZ-VSL~\cite{mo2022EZVSL} for localization, and Sound-of-Pixels~\cite{zhao2018the} and MP-Net~\cite{xu2019mpnet} for source separation. These comparisons again demonstrate the added functionality and improved performance of a unified framework like \method.

\subsection{Experimental analysis}

In this section, we present the results of our ablation studies aimed at assessing the effectiveness of the various components of \method on audio-visual source separation, localization, and recognition. 
Specifically, we investigate the impact of Correspondence \& Localization (CL), Mixed Audio Separation (MAS), and Mixed Visual Alignment (MVA) on the performance of our approach. We also analyze the impact of the decoder depth for MAS, and the mixture ratio $\alpha$ in MVA, to provide insights into the optimal \method configuration. All experiments were conducted on the VGGSound-MUSIC dataset.

\noindent\textbf{CL, MAS, and MVA objectives.}
We assessed the effectiveness of each objective on the method's performance.
Table~\ref{tab:ab_module} presents the model's performance on the source localization, separation, and recognition tasks.
As can be seen, both source localization and recognition tasks can be significantly enhanced by adding the Mixed Audio Separation and Mixed Visual Alignment objectives. 
Specifically, the full \method outperforms a model trained with Correspondence \& Localization (CL) alone by 4.3 PIAP, 9.4 Precision, and 13.5 F1 score on the localization task (Table~\ref{tab:vsl}), and by 12.5 IR-Acc, 6.0 xNN-Acc and 1.4 wNN-Acc on nearest neighbor recognition/retrieval (Table~\ref{tab:avrec}). 
Sound source separation can also be enhanced significantly by adding the Correspondence \& Localization, and Mixed Visual Alignment objectives, yielding a gain of 1.53 SDR and 1.6 SAR over a method trained for separation alone (Table~\ref{tab:avss}). 

These results show that localization helps separation and vice-versa, highlighting the strong interdependence of the two tasks and underscoring the importance of jointly optimizing them in our proposed audio-visual learning framework.
It also shows that the proposed MVA regularization can help all tasks significantly and thus is a valuable addition to the proposed unified framework.

\begin{figure}[t]
\centering
\begin{subfigure}{0.49\linewidth}
    \centering
    \includegraphics[width=0.96\linewidth]{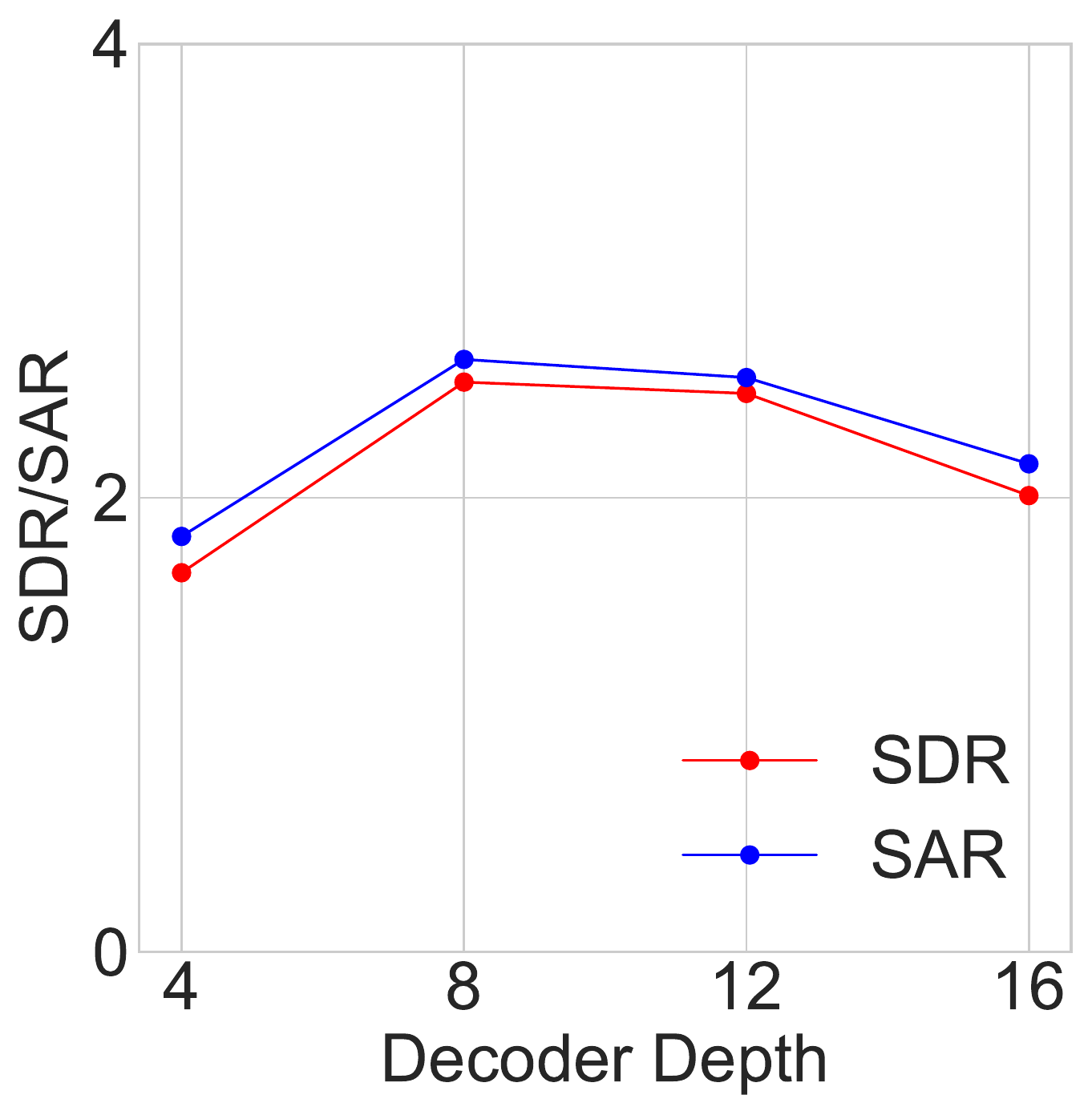}
\end{subfigure}%
\begin{subfigure}{0.49\linewidth}
    \centering
    \includegraphics[width=\linewidth]{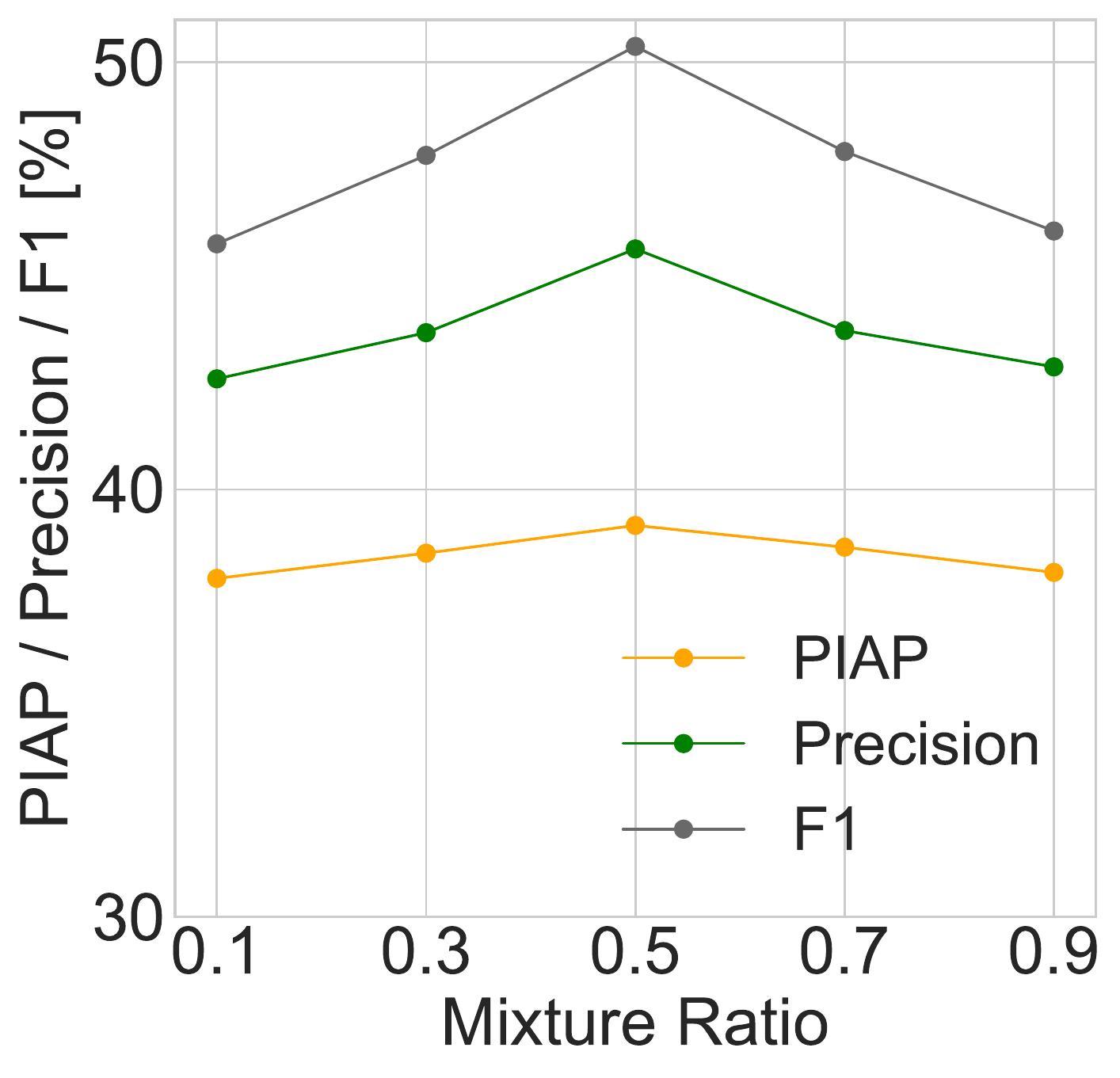}
\end{subfigure}
\caption{Impact of decoder depth in MAS on source separation and mixture ratio in MVA on source localization. \label{fig: ab_depth_alpha}
}
\end{figure}

\noindent\textbf{Decoder depth in MAS.}
The depth of the decoder used for source separation can affect separation performance.
To assess the impact of the decoder depth, we varied it from $\{4, 8, 12, 16\}$. As shown in Figure~\ref{fig: ab_depth_alpha}, the optimal decoder depth is 8, achieving the best separation both in terms of SDR and SAR. Increasing the decoder depth beyond eight hurt performance due to overfitting.

\noindent\textbf{Mixture ratio in MVA.}
Mixed visual alignment (MVA) can help with learning separable visual features that are aligned with multiple sound sources. 
The mixture ratio $\alpha$ is a critical hyper-parameter of MVA and can after performance significantly.
To better understand the effect of the mixture ratio, we show the localization performance for varying ratios in Figure~\ref{fig: ab_depth_alpha}.
\method obtained optimal performance at a mixture ratio of 0.5, according to all metrics.

\begin{table}[!t]
    \centering
    \caption{{\bf Localization with multiple sound sources.} Quantitative results of Precision on VGGSound-All dataset.}
    \label{tab: exp_num_src}
    \scalebox{0.9}{
    \begin{tabular}{lcccccc}
        \toprule
        Method & 1 & 2 & 3+ \\	
        \midrule
        CCoL & 35.31 & 32.25 & 30.51 \\
        EZ-VSL & 37.79 & 31.67 & 29.32 \\
        \method (ours) & \textbf{39.68} & \textbf{35.59} &  \textbf{33.76} \\
    \bottomrule
    \end{tabular}}
\end{table}

\noindent\textbf{Localization with multiple sound sources.}
Lastly, we analyzed the source localization performance on samples containing multiple sound sources. 
Table~\ref{tab: exp_num_src} compares three methods, CCoL, EZ-VSL, and our \method, for localization on the VGGSound-All dataset. 
EZ-VSL focuses on localization (without separation), CCoL performs joint localization and separation, and our method further trains for mixed visual alignment and global audiovisual correspondence (in addition to localization and separation). 
As can be seen, EZ-VSL is better than CCoL with a single object (\ie when no separation is needed), but its performance drops faster as the number of objects increases.
This result indicates that adding a separation objective helps localization, especially for samples with a larger number of objects. 
The proposed approach, \method, not only outperforms EZ-VSL for videos with a single sound source, but also shows a slower (although still noticeable) decay of performance as the number of active sound sources increases.

\subsection{Limitations}
Although \method achieves superior results on all three audio-visual downstream tasks (\ie source localization, separation, and recognition), the performance gains over prior work on source separation are less consistent than those on localization. For example, our method achieves lower SAR on the MUSIC dataset than other recent methods like MP-Net. We highlight, however, both the added functionality of the model (\ie, its ability to simultaneously address three tasks, as opposed to a single task) and performance improvements on other tasks when separation is used as a learning objective.

\section{Conclusion}
In this work, we present \method, a simple yet effective approach that unifies audio-visual learning for different tasks, including localization, separation, and recognition.
Specifically, we leverage correspondence and localization to align the representations of corresponding audio and video frames. 
We also use a mixed audio separation objective to capture discriminative audio representations from mixed audio, and introduce a mixed visual alignment objective to learn separable visual features from mixed images that can be aligned with individual sound sources.
Through extensive experiments on MUSIC, VGG-Instruments, VGG-Music, and VGGSound-All datasets, we demonstrate the effectiveness of all components of our \method framework and achieve favorable results in comparison to prior work on the tasks of visual sound source localization, separation, and nearest neighbor recognition. 

\noindent\textbf{Broader Impact.}
The proposed method unifies sound source localization, sound separation, and recognition from user-uploaded web videos, which might cause the model to learn internal biases in the data.
For example, the model could fail to localize, separate, and recognize certain rare but crucial sound sources.
These issues should be carefully addressed when it comes to the deployment of real scenarios.

{\small
\bibliographystyle{icml2023}
\bibliography{reference}
}

\end{document}